\documentclass[prd,aps,twocolumn,floats,floatfix,nofootinbib,superscriptaddress] {revtex4-1}
\usepackage{graphicx}
\usepackage{dcolumn}
\usepackage{bm}
\usepackage{graphics}
\usepackage{slashed}
\usepackage{amssymb}
\usepackage{natbib}
\usepackage{amsmath}
\usepackage{url}
\usepackage[usenames,dvipsnames,svgnames,table]{xcolor}
\usepackage{color}
\usepackage{xcolor}
\usepackage{tabularx}
\usepackage{hyperref}
\usepackage{enumitem}
\usepackage{soul}
\usepackage{grffile}
\usepackage{algorithm}
\usepackage[noend]{algpseudocode}
\usepackage{mathtools}
\usepackage[normalem]{ulem}   

\newcommand{\trT}{\mathrm{tr}T}

\def\beq{\begin{equation}}
\def\eeq{\end{equation}}
\newcommand{\bea}{\begin{eqnarray}}
\newcommand{\eea}{\end{eqnarray}}
\newcommand{\mhduet}{{\texttt{MHDuet}~}} 
\newcommand{\mhduetX}{{\texttt{MHDuet}}} 
\newcommand{\FUKA}{{\texttt{FUKA}~}} 
\hypersetup{
	colorlinks=true,
	linkcolor=red,
	linkbordercolor=black,
	citecolor=blue,
	filecolor=magenta,      
	urlcolor=magenta,
}

\begin{document}

\title{Implications of Magnetic Flux--Disk Mass Correlation in Black Hole--Neutron Star Mergers for GRB sub-populations}

\author{Manuel R. Izquierdo}
\address{Departament  de  F\'{\i}sica $\&$ IAC3,  Universitat  de  les  Illes  Balears,  Palma  de  Mallorca,  Baleares  E-07122,  Spain}

\author{Carlos Palenzuela}
\address{Departament  de  F\'{\i}sica $\&$ IAC3,  Universitat  de  les  Illes  Balears,  Palma  de  Mallorca,  Baleares  E-07122,  Spain}

\author{Steven Liebling}
\address{Long Island University, Brookville, New York 11548, USA}	

\author{Ore Gottlieb}
\address{Center for Computational Astrophysics, Flatiron Institute, New York, NY 10010, USA}
\address{Department of Physics and Columbia Astrophysics Laboratory,	Columbia University, Pupin Hall, New York, NY 10027, USA}	

\author{Miguel Bezares}
\address{Nottingham Centre of Gravity,
	Nottingham NG7 2RD, United Kingdom}
\address{School of Mathematical Sciences, University of Nottingham,
	University Park, Nottingham NG7 2RD, United Kingdom}
\begin{abstract}
We perform numerical relativity simulations of black hole--neutron star (BH--NS) mergers with a fixed mass ratio of $q = 3$, varying the BH spin to produce a wide range of post-merger accretion disk masses. 
Our high-order numerical scheme, fine resolution, and Large Eddy Simulation techniques enable us to achieve 
likely the most resolved BH--NS merger simulations to date, capturing the  post-merger magnetic field amplification 
driven by turbulent dynamo processes.
Following tidal disruption and during disk formation, the Kelvin-Helmholtz instability in the spiral arm drives a turbulent state in which the magnetic field, initialized to a
realistic average value of $10^{11}\, \rm{G}$, grows to an average of approximately $10^{14}\, \rm{G}$ in the first $\approx 20\, \mathrm{ms}$ post-merger.
Notably, the dimensionless magnetic flux on the BH, $ \phi $, evolves similarly across nearly two orders of magnitude in disk mass. This similarity, along with estimates from longer numerical simulations of the decay of the mass accretion rate,  suggests a universal timescale at which the dimensionless flux saturates at a magnetically arrested state~(MAD) such that $ \phi \approx 50 $ at $t_{\rm MAD} \gtrsim 10\,{\rm s}$.
The unified framework of Gottlieb et al. \cite{Gottlieb2023} established that the MAD timescale sets the duration of the resulting compact binary gamma-ray burst (cbGRB), implying that all BH--NS mergers contribute to the recently detected new class of long-duration cbGRBs.
\end{abstract} 

\maketitle

\section{Introduction}
Mergers of a black hole~(BH) with a neutron star~(NS) are among the most fascinating and energetic events in the universe, providing exceptional chances to study extreme physics. The Ligo-Virgo-Kagra (LVK) collaboration has detected three such mergers, none of which is associated with any observed electromagnetic counterpart~\cite{LIGOScientific:2021qlt,KAGRA:2021vkt,Gupta:2023evt,Zhu:2021jbw}. In fact, the most recent of these three events, GW230529, stands out as particularly intriguing as its remnant appears to lie within a potential mass gap~\cite{LIGOScientific:2024elc}. Although BH--NS mergers are captivating gravitational wave sources, a subset of such mergers in the appropriate parameter regime is expected to produce electromagnetic signals. With the planned improvement of the sensitivities of LVK detectors, the hope is that gravitational observations of BH--NS mergers will become routine and that accompanying electromagnetic signals, whether a jet-powered gamma-ray burst (GRB) or a kilonova, will be found.

One of the most critical questions for inferring GRB physics from BH--NS and NS--NS mergers is how the magnetic flux on the post-merger BH scales with the disk mass. Ref.~\cite{Gottlieb2023} developed a theoretical framework demonstrating that the time of flux saturation on the BH determines the GRB duration. Namely, the amount of flux indicates whether the resulting GRB is a standard short-duration binary GRB (sbGRB) or is of the new long-duration binary GRB (lbGRB) class \citep[see, e.g.,][]{Rastinejad2022}. They further demonstrated that if the magnetic flux scales quasi-linearly with the disk mass, BHs can generate both lbGRBs and sbGRBs. In contrast, if the magnetic flux is quasi-universal across different disk masses, as might be expected when the field is generated through turbulent dynamo, all BH-powered jets produce lbGRBs.

Recently, Ref.~\cite{Gottlieb2024} demonstrated that the color and brightness dependency in GRB-associated kilonova observations points to NSs as the central engines of sbGRBs. 
If NS--NS mergers are indeed the engines of all sbGRBs, it follows that BH-powered jets drive lbGRBs. The persistence of all BH-powered jets for $ \gtrsim 2\,{\rm s} $ suggests a similar evolution of the dimensionless flux across a range of disk masses.

However, this inference has yet to be tested through a systematic suite of numerical simulations. In this study, we conduct a series of numerical relativity simulations of BH--NS mergers to investigate the dependence of the magnetic field on the disk mass. In particular, we pay special attention to the magnetic flux on the BH as a function of the disk mass and its implications for
the GRB class.

While previous numerical relativity studies have explored various physical aspects of BH--NS mergers, such as gravitational waves, disk formation and mass ejection, eccentric binaries, different mass ratios or spins,  finite temperature and neutrino transport, different equations of state for the NS, and  post-merger evolution ~\cite{Loffler:2006wa,Shibata:2006ks,Etienne:2007jg,Duez:2008rb,Etienne:2008re,Etienne:2011ea,Foucart:2012nc,Foucart:2014nda,Paschalidis:2014qra,Kiuchi:2014hja,Kyutoku:2015gda,East:2015yea,Foucart:2016vxd,Foucart:2018rjc,Ruiz:2020elr,Chaurasia:2021zgt,Most:2021ytn,Hayashi:2022cdq,Hayashi:2022cdq,Martineau:2024zur,Chen:2024ogz,Matur:2024nwi,Topolski:2024jva,Topolski:2024beu,Kim:2024fuy}~(while comprehensive overviews of many simulations can be found in the reviews~\cite{Foucart:2020ats,Kyutoku:2021icp,Duez:2024rnv}), this work extends the study started in~\cite{Izquierdo:2024rbb} that focused on a comprehensive analysis of the amplification mechanism of the magnetic field in BH--NS mergers.
In particular, we considered a system in which the NS is disrupted before reaching the innermost stable circular orbit, forming a massive torus-like accretion disk.  Just before the disruption phase, the magnetic field strength was set to $10^{11} \text{G}$, an upper bound of the values observed in old NSs, but still several orders of magnitude lower than typical values commonly used in numerical studies~\cite{Etienne:2011ea,Kiuchi:2015qua,Hayashi:2022cdq,2022PhRvD.106b3008H}. Using high-order numerical schemes, Large Eddy Simulation techniques, and a spatial resolution of ${\cal O}(100-200)\,\mathrm{m}$, we found that a quasi-stationary accretion disk is quickly formed in $\approx 10\,\rm{ms}$, with the magnetic field saturating at $B\approx 10^{14}\,\mathrm{G}$.

Here, we present a follow-up on Ref.~\cite{Izquierdo:2024rbb} to address the dependence of the magnetic field on the disk mass. As in the previous work, we set a realistic initial magnetic field in the NS and employ the same numerical techniques with an even higher resolution of $90\,\mathrm{m}$. To the best of our knowledge, these are the highest-resolution simulations of BH--NS mergers performed to date. 
Our approach highlights key differences in the post-merger dynamics compared with other works that set a much stronger initial magnetic field in the NS. Initializing the simulations with an unrealistically high magnetic field might generally lead to several nonphysical effects. First, the magnetic energy soon after disk formation might already be comparable to the fluid's kinetic energy, suppressing the effect of the Kelvin-Helmholtz instability (KHI). As a result, the generated magnetic field lacks the small-scale, isotropic structure arising from turbulent dynamo amplification, as seen in NS--NS simulations that resolve the KHI~\cite{aguilera2020,Palenzuela:2021gdo,Aguilera-Miret:2021fre, Palenzuela:2022kqk, 2023PhRvD.108j3001A,Aguilera-Miret:2024cor}. Instead, the field maintains a larger-scale structure that might facilitate the magneto-rotational instability (MRI) to be active already at very early times. The formation of a poloidal magnetic field and jets (i.e., see for instance~\cite{Paschalidis:2014qra,Hayashi:2024jwt}), occurs under conditions that might be quite different from the chaotic and turbulent magnetic environment expected with physically realistic initial conditions. 
Furthermore, we present evidence that, with the numerical techniques and high resolution employed, the saturation strength of the magnetic field is converging.

This paper is organized as follows. In Sec.~\ref{sec:setup}, we briefly summarize the method and initial setup employed for the numerical simulations. Following this, in Sec.~\ref{sec:results}, we present our numerical results from the BH--NS merger. Finally, we conclude with a discussion in Sec.~\ref{sec:discussion}. We follow the usual convention according to which indices are written as $\{a,b,c,d\}$ $(\{i,j, k\})$ refer to spacetime (spatial) coordinates and adopt a metric signature $(-,+,+,+)$.

\section{Setup}
\label{sec:setup}
Here, we provide an overview of the equations of motion, the initial data, and the numerical setup employed in this work. The numerical simulations carried out in this paper have been performed using our publicly available \mhduet code \cite{mhduet_webpage}. Previously, it has been utilized to simulate the coalescence of NS--NS using LES techniques to study the magnetic field amplification mechanisms~\cite{aguilera2020,Palenzuela:2021gdo,Aguilera-Miret:2021fre, Palenzuela:2022kqk, 2023PhRvD.108j3001A,Aguilera-Miret:2024cor} and the effects of phase transitions occurring during the merger~\cite{2021CQGra..38k5007L}.
More details on the formulation of the Einstein equations can be found in Ref.~\cite{bezares17} and the numerical methods in Refs.~\cite{simf3,liebling20}.
Recently we have incorporated some new features to deal with binary BH--NS mergers~\cite{Izquierdo:2024rbb}: the ability to evolve accurate initial data provided by the \FUKA library~\cite{Papenfort:2021hod,GRANDCLEMENT20103334},
a slight modification of the gauge conditions, and a new treatment to address the  magnetized fluid as it falls within the BH horizon.
%

\subsection{Evolution formalism}
Here we adopt the Covariant Conformal Z4 (CCZ4) formulation to describe the spacetime geometry. The Z4 is an extension of Einstein's equations that includes a four-vector $Z^a$ to measure the deviation from Einstein's solutions~\cite{bona03,alic,bezpalen}, i.e.,  
\begin{eqnarray}
R_{ab} &+& \nabla_a Z_b + \nabla_b Z_a   = 
8\pi \, \left( T_{ab} - \frac{1}{2}g_{ab} \,\trT \right) \nonumber \\
&+& \kappa_{z}  \left(  n_a Z_b + n_b Z_a - g_{ab} n^c Z_c \right)~,
\label{Z4cov}
\end{eqnarray}
where $T_{ab}$ is the stress-energy tensor describing the matter content. Additional terms, proportional to $\kappa_{z}$, are added to enforce the dynamical damping of the energy and momentum constraints associated with $Z^{a}$~\cite{gundlach}.

The 3+1 decomposition splits the four-dimensional tensors and equations into space and time components, such that the covariant equations can be written as a time evolution system. The line element is given by
\begin{equation}
ds^2 = - \alpha^2 \, dt^2 + \gamma_{ij} \bigl( dx^i + \beta^i dt \bigr) \bigl( dx^j + \beta^j dt \bigr)~, 
\label{3+1decom}  
\end{equation}
where $\alpha$ is the lapse function, $\beta^{i}$ is the shift vector, and $\gamma_{ij}$ is the induced metric on each spatial foliation. In the CCZ4 formulation, a conformal transformation is performed of the evolved fields (i.e., the metric and extrinsic curvature) in order to achieve robust and stable simulations in black-hole spacetimes. The additional constraints associated with this change of evolved fields are also enforced dynamically with damping terms~\cite{bezpalen}. Finally, the evolution system is supplemented with suitable gauge conditions, namely the Bona-Mass\'o~\cite{BM} slicing and the Gamma-driver shift~\cite{alcub} condition, which ensure that the final evolution system is strongly hyperbolic.
   
The magnetized fluid is modeled using the general relativistic magnetohydrodynamics~(GRMHD) equations in flux-conservative form as described for instance in Ref.~\cite{palenzuela15}. We evolve the standard conservative fields, $\left\lbrace D, S^i , U, B^i \right\rbrace$, which are functions of the primitive ones $\left\lbrace\rho,\epsilon,v^{i},B^i\right\rbrace$. Here $\rho$ is the rest-mass density of the fluid, $\epsilon$ its specific internal energy, $p$ its pressure, $v^{i}$ is the three-velocity vector measured by Eulerian observers and $B^i$  represents the magnetic field vector. The primitive and the conserved fields are related by the non-linear relations
\begin{eqnarray}
D &=& \rho W ~, \\
S^i &=& (h W^2 + B^2 ) v^i - (B^k v_k) B^i ~,\\
U &=& h W^2 - p + B^2 - \frac{1}{2} \left[ (B^k v_k)^2 + \frac{B^2}{W^2} \right]  ~,
\end{eqnarray}
where $h=\rho(1+\epsilon)+p$ is the enthalpy and $W=1/\sqrt{1-v_{i}v^{i}}$ is the Lorentz factor. A closure relation is required to recover the primitive variables from the conserved fields, known as the equation of state (EoS) that expresses the pressure as a function of the other thermodynamic variables $p=p(\rho, \epsilon)$. We enforce dynamically the solenoidal constraint on the magnetic field using the hyperbolic divergence cleaning approach~\citep{palenzuela18}.

The GRMHD equations are augmented by sub-grid-scale (SGS) terms which appear in the context of LES techniques. The main idea of LES is to enhance the effective resolution of the simulation by accounting for the effect of unresolved small-scales into the large scales through the gradient SGS model~\cite{vigano20,aguilera2020}. These new terms help capture the turbulent phenomena appearing after the merger that are crucial to describe accurately the magnetic field amplification. For all our simulations, the LES terms are applied only above a threshold density $\rho \geq 6\times 10^{8} \text{g/cm}^{3}$ in order to avoid nonphysical effects occurring in rarefied regions.
Finally, smooth dissipation is applied well inside the apparent horizon on the fluid fields to prevent numerical instabilities from occurring due to the failure of the inversion from conserved to  primitive fields. Further details on the evolution equations and the parameters used in the simulations can be found in Ref.~\cite{Izquierdo:2024rbb}.

\subsection{Initial Data}\label{ID}

We build initial data consisting of a BH and a NS in a quasi-circular orbit using the public library \FUKA \cite{Papenfort:2021hod,GRANDCLEMENT20103334}. \FUKA is an initial data solver using the eXtended Conformal Thin-Sandwich formulation of Einstein's field equations for various compact object configurations, including spinning BH--NS binaries. The neutron star is modeled employing a piecewise polytropic EoS representing the cold part of the APR4 EoS.

We explore binaries that form a non-negligible  accretion disk $M_{\rm{disk}} \approx 0.001-0.1\,M_{\odot}$ around the final black hole. According to Table~III of Ref.~\cite{Kyutoku_2015}, with the soft EoS APR4 and mass ratio $q=M_{\rm BH}/M_{\rm NS}=3$, the resulting disk mass  is only $M_\mathrm{disk} \approx 4 \times 10^{-4}\,M_\odot$ for non-spinning black holes but increases significantly to $M_\mathrm{disk} \approx 0.08\,M_\odot$ when the black hole spin is $a/M_{\rm{BH}}=0.5$. Since we aim to study a wide range of disk masses, we construct initial data corresponding to this setup. The mass of the black hole is $M_\mathrm{BH}=4.05\,M_{\odot}$, while we consider three different spins $a/M_{\rm{BH}}=(0.2,0.35,0.5)$. The mass of the neutron star $M_\mathrm{NS}=1.35\, M_{\odot}$ and its  radius is $R_\mathrm{NS} = 11.2\,\rm{km}$. The binary is initially in a quasi-circular orbit with a separation of $44\,\rm{km}$. With this information one can already estimate the radius of the innermost stable circular orbit (ISCO), $R_\mathrm{ISCO}$, which is the main quantity that determines the amount of mass that will form the disk after the star's disruption. Using the method described in Ref.~\cite{Hanna_2008} one can estimate the final spin as well as $R_\mathrm{ISCO}$, obtaining that $R_\mathrm{ISCO}/M=(3.73,3.45,3.16)$ respectively for the three spins $a/M_{\rm{BH}}=(0.2,0.35,0.5)$.

At half an orbit before the merger, we set a purely poloidal magnetic field confined to the interior of the neutron star. The average magnetic field strength is $10^{11} \text{G}$, consistent with the observed upper range of old neutron stars in a binary system, but several orders of magnitude less than the large initial fields used in other simulations (e.g., \cite{Etienne:2011ea,Kiuchi:2015qua,Hayashi:2022cdq,2022PhRvD.106b3008H}). The same magnitude of the magnetic field was considered in our previous work \cite{Izquierdo:2024rbb}, in which we first explored the case $a/M_{\rm{BH}}=0.5$.

\subsection{Numerical methods}

We use our code \mhduetX, generated automatically by the platform {\sc Simflowny} \cite{arbona13,arbona18} to run under the {\sc SAMRAI} infrastructure \cite{hornung02,gunney16}, which provides the parallelization and the adaptive mesh refinement. The code, which has been extensively tested in several scenarios \cite{palenzuela18,vigano19,vigano20,liebling20}, employs the following numerical methods: fourth-order-accurate operators for the spatial derivatives in the SGS terms and in the Einstein equations (the latter are supplemented with sixth-order Kreiss-Oliger dissipation); a high-resolution-shock-capture (HRSC) method for the fluid, based on the  Lax-Friedrich flux splitting formula \cite{shu98} and the fifth-order reconstruction method MP5 \cite{suresh97}; a fourth-order Runge-Kutta scheme with sufficiently small time step $\Delta t \leq 0.4\,\Delta x$; and an efficient and accurate treatment of the refinement boundaries when sub-cycling in time~\cite{McCorquodale:2011,Mongwane:2015}. A complete assessment of the implemented numerical methods can be found in Refs.~\cite{palenzuela18,vigano19}.

The BH--NS binaries considered here are evolved in a cubic domain of size $\left[-1152\, \mathrm{km},1152\, \mathrm{km}\right]^3$. During the inspiral, the binary is fully covered by the coarse grid, $6$ fixed mesh refinement~(FMR) and 1 adaptive mesh refinement~(AMR) levels,  reaching a minimum grid spacing $\Delta x=90\, \rm{m}$ within the NS. Each refinement level consists of a rectangular cuboid with twice the resolution of the next larger one. During the tidal disruption and disk formation, we change the refinement grid structure and keep it fixed afterwards, providing a uniform resolution throughout the one-arm spiral and the shear layers. In particular, for all our runs, both the black hole and the densest region of the remnant disk are contained within the finest refinement level covering a  box $[-52, 52][-52, 52][-22, 22]\mathrm{km}^3$ with resolution $\Delta x=90\, \rm{m}$. Note that this is, to the best of our knowledge, the highest resolution employed in this kind of BH--NS simulation.

\subsection{Analysis quantities}

Here we summarize the analysis quantities that we use to monitor the dynamics in our simulations. Most of them were described in detail in Ref.~\cite{Palenzuela:2021gdo} and were also employed in~\cite{Izquierdo:2024rbb}. Thus, here we focus on the main quantities and the ones that have not been presented before.

We compute global quantities, integrated over the whole computational domain, such as the total baryonic mass $E_{\rm{bar}}$, magnetic energy $E_{\rm{mag}}$, thermal energy $E_{\rm{th}}$, and rotational kinetic energy $E_{\rm{rot}}$. The averages for a given quantity $\mathcal{Q}$ over a certain region are as follows: $\langle \mathcal{Q} \rangle$ when it is averaged over the entire disk, and $\langle \mathcal{Q} \rangle_\mathrm{bulk}$ when it is averaged over the bulk. Here we consider \textit{disk} as any fluid region with densities above $10^8 \rm{g/cm}^3$, and \textit{bulk} as the densest part of the disk containing $75\%$ of its total mass. Some of these quantities include the fluid angular velocity $\Omega = \frac{u^{\phi}}{u^t}$, the inverse of the plasma beta parameter $\beta^{-1} \equiv \frac{B^2}{2 p}$, and the wavelength of the fastest growing mode of the axisymmetric MRI,  $\lambda_{\textrm{MRI}}=(2 \pi/\Omega) (B_z/\sqrt{4 \pi h + B^2})$.

We also compute fluxes of certain quantities over spherical surfaces at different radii. In particular, we compute the flux of mass passing through a sphere of radius $r$ as
\begin{equation}
	\dot M=\int \rho W (\alpha v^r -\beta^r) r^2 \sqrt{\gamma} d\Omega~,
\end{equation}
where $d\Omega$ is the differential of solid angle. The mass accretion rate $\dot M_\mathrm{acr}$ is the flux of mass passing through the spherical surface at $R \approx 9\,$km, larger than the black hole´s apparent horizon but well inside the ISCO.
We also compute the magnetic flux $\Phi_B$ and the dimensionless flux $\phi$, defined as:
\begin{eqnarray}
	\Phi_B =\int |B^r| r^2 \sqrt{\gamma} d\Omega ~~~~,~~~~	
    \phi = \frac{\Phi_B}{\sqrt{r^2 \dot M}}~,
\end{eqnarray}
which we evaluate also on a spherical surface slightly beyond the apparent horizon.

Further information can be obtained from the distribution of the kinetic and magnetic energies over the spatial scales (i.e., the spectra), which is computed on a domain containing the most significant part of the disk remnant. We take the Fourier transform of the solution within a cubic box $[-66\, \text{km} ,66\, \text{km}]^3$ with a resolution equal to that of the finest in each simulation, interpolating from coarser grids if needed in order to have a unigrid box.
Further details of the numerical procedure to calculate the spectra, and in particular the energy per wavenumber ${\cal E}(k)$, can be found in~\cite{aguilera2020,vigano19,vigano20}.
With these spectrum distributions we can define the spectrum-weighted average wavenumber and its associated length scale 
\begin{equation}
	\langle k \rangle \equiv \frac{\int_k k\,{\cal E}(k) \,dk} {\int_k {\cal E}(k)\, dk}~~~,~~~
	\langle L\rangle = 2\pi/\langle k\rangle~,
	\label{eq:coherence}
\end{equation}
which represents the typical coherence scale of the structures present in the field.

\begin{table*}[ht]
	\centering
	\renewcommand{\arraystretch}{1.3} 
		\begin{tabular}{l||ccccccccccccc}
			\toprule			
			\textbf{$a/M_{\rm{BH}}$} & $J/M^2$ & $M~[M_{\odot}]$ & $M_{\rm{disk}}~[M_{\odot}]$  & $M_{\rm{eje}}~[M_{\odot}]$ &  \textbf{$\dot{M}_{\rm{acr}}~[M_{\odot}/\mathrm{ms}]$} &\textbf{$E_{\rm{rot}}~[\mathrm{ergs}]$} & \textbf{$E_{\rm{th}}~[\mathrm{ergs}]$} & \textbf{$E_{\rm{mag}}~[\mathrm{ergs}]$} & \textbf{$\langle B \rangle_\mathrm{bulk}~[G]$ } & \textbf{$\langle \beta^{-1} \rangle_\mathrm{bulk}$} &\textbf{$\phi_\mathrm{BH}$}  \\
            \toprule
			$0.2$  & $0.594$ & $5.37$ &$9.4\times 10^{-4}$ & $9.3\times 10^{-4}$ & $1.2\times 10^{-4}$ & $1.5\times 10^{50}$ & $2.6\times 10^{49}$ & $ 5.0\times 10^{47}$ & $7.8\times 10^{13}$ & $0.026$ & $0.75$ \\
			$0.35$ & $0.677$ & $5.32$ & $7.7\times 10^{-3}$ & $2.3\times 10^{-3}$ & $9.0\times 10^{-4}$ & $1.3\times 10^{51}$ & $1.7\times 10^{50}$ & $1.6\times 10^{48}$ & $1.5\times 10^{14}$ & $0.017$ & $0.58$ \\
			$0.5$  & $0.742$ & $5.29$ & $4.2 \times 10^{-2}$  & $7.3\times 10^{-3}$ & $1.4\times 10^{-3}$ & $6.3\times 10^{51}$ & $7.2\times 10^{50}$ & $4.9\times 10^{48}$ & $2.2\times 10^{14}$ & $0.008$ & $0.57$ \\
 			\hline 
		\end{tabular}
		\caption{\textit{Summary of simulation results}. Several relevant quantities, evaluated at $t = 20$~ms after the merger, for the different dimensionless spins $a/M_{\rm{BH}}$: the dimensionless final spin  $J/M^2$ with $J$ the total angular momentum and $M$ the total mass of the system, the disk's mass $M_{\rm{disk}}$, the ejected mass $M_{\rm{eje}}$, the mass accretion rate in the BH $\dot{M}_{\rm{acr}}$, the rotational kinetic energy $E_{\rm{rot}}$, the thermal energy $E_{\rm{th}}$, the magnetic energy $E_{\rm{mag}}$, the magnetic field $\langle B \rangle_\mathrm{bulk}$ and the inverse plasma beta $\langle \beta^{-1} \rangle_\mathrm{bulk}$ averaged in the bulk, and the dimensionless magnetic flux $\phi$ evaluated at the BH horizon. \label{tab:spin_results}}
\end{table*}

\begin{figure*}[!ht]
	\includegraphics[width=18cm]{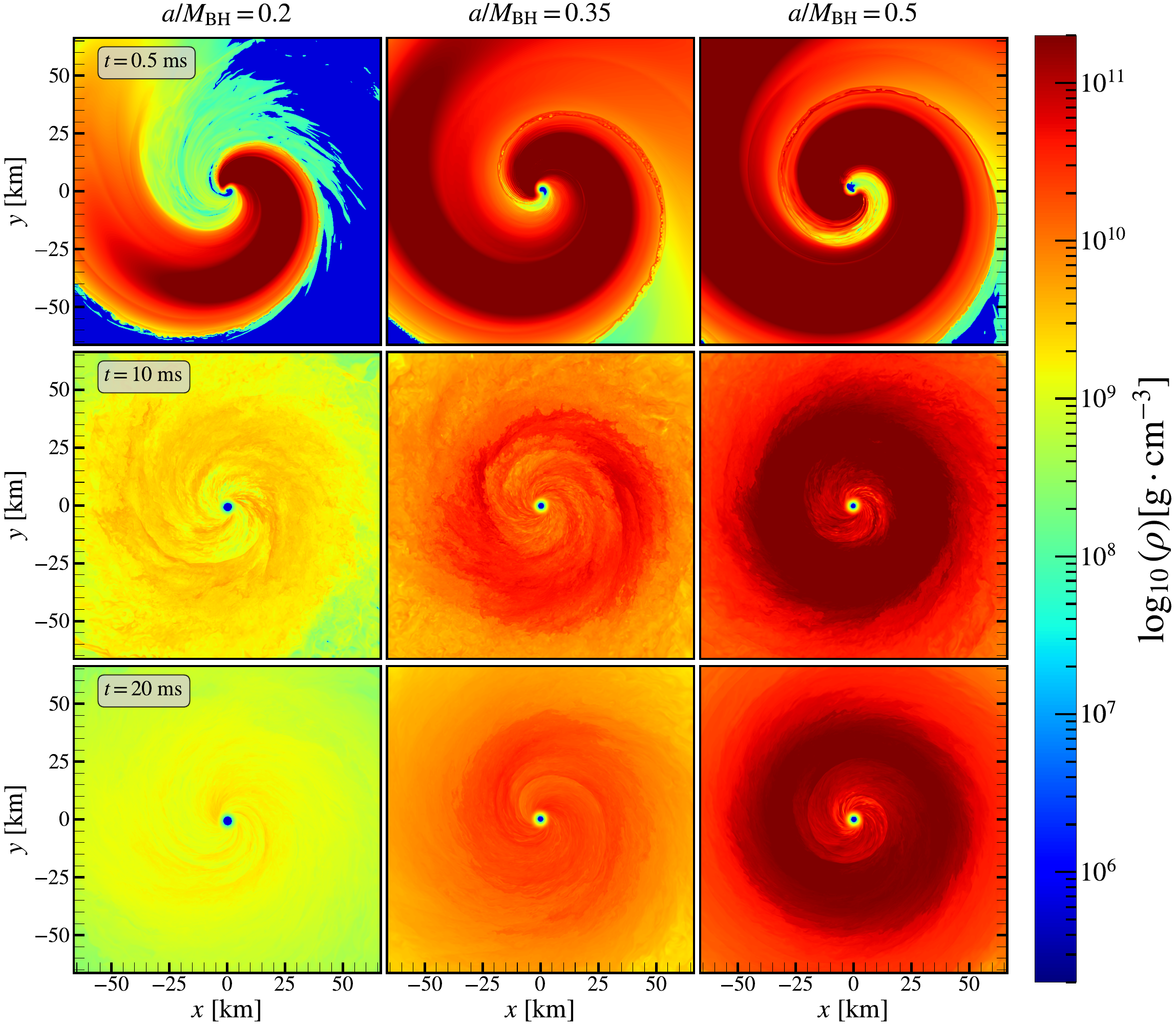}
	\caption{\emph{Evolution of the density in the orbital plane}. Snapshots of the rest-mass density at times $t = (0.5, 10, 20)$ ms \textbf{(from top to bottom)} for the three configurations corresponding to BH spin $a/M_{\rm{BH}} = (0.2, 0.35, 0.5)$ \textbf{(from left to right)}. The NS is tidally disrupted, forming a one-arm spiral structure that winds around the BH. Shear layers are present when the head of this structure collides with its own tail and forms a turbulent and rapidly rotating torus, which relaxes to a roughly axisymmetric disk after approximately $20$~ms. The qualitative dynamics is similar irrespectively of the spin, although the mass of the final disk strongly depends on the BH's spin. \label{RHOF}}
\end{figure*}

\begin{figure*}[p]
	\centering
	\includegraphics[width=\textwidth,height=\textheight,keepaspectratio]{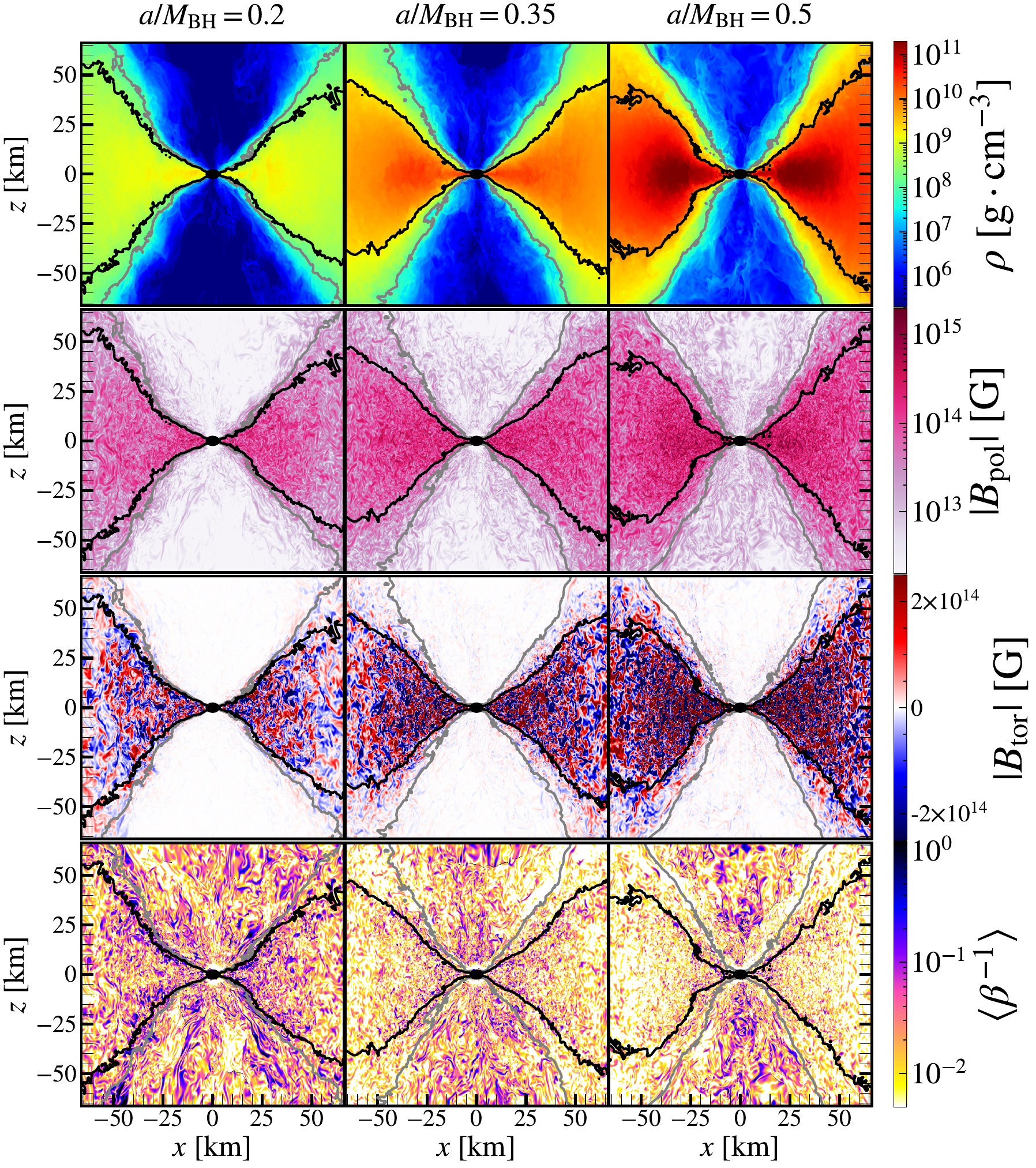}
	\caption{\emph{Several quantities in the meridional plane of the final disk}.
		Snapshots of the rest-mass density \textbf{(top)}, the poloidal and toroidal magnetic field components \textbf{(middle)}, and the inverse of $\beta$ \textbf{(bottom)} at $t = 20\, \text{ms}$ after the merger for $a/M_{\rm{BH}}=(0.2,0.35,0.5)$ \textbf{(left, middle and right, respectively)}. The grey and black contour-lines indicate the separatrix between funnel, envelope and bulk for each case. Although both the density and magnetic field are larger for higher spins, $\beta^{-1}$ is more relevant for small spins. Notice the comparable values of the poloidal and toroidal magnetic field as well as its turbulent dynamics.
	}
	\label{meridional}
\end{figure*}

\section{Results}
\label{sec:results}

In this section, we present the results of our binary BH--NS simulations. First, we describe  qualitatively the dynamics focusing on the evolution of the magnetic field. Then, we study various global quantities and fluxes at the black hole's horizon, the radial profile of different physical fields in the accretion disk, and finally the energy spectra. 

\subsection{Qualitative dynamics}

\begin{figure}[ht!]
	\includegraphics[width=8.5cm]{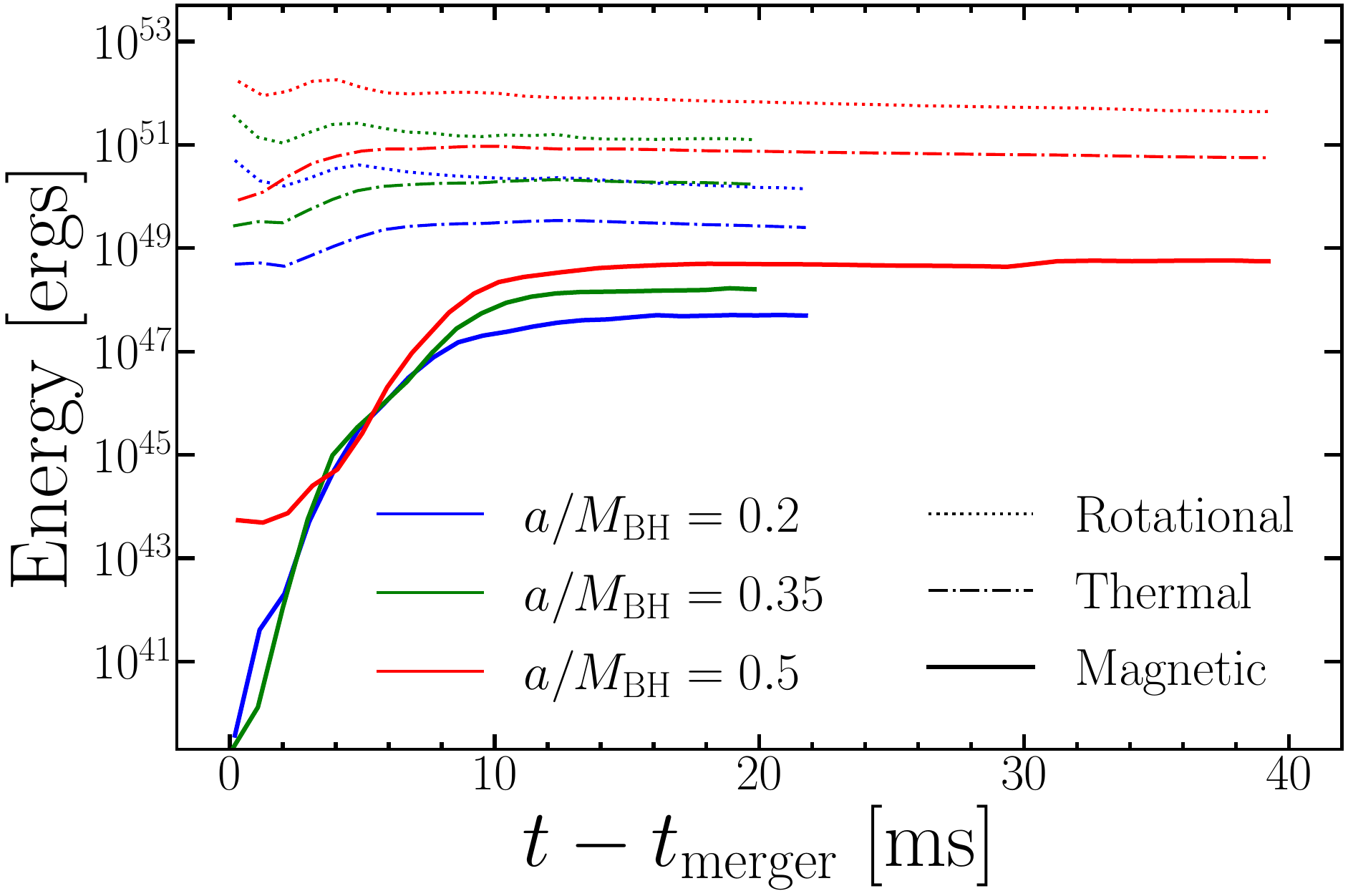}
	\caption{\emph{ Energies evolution.}
		Rotational \textbf{(dotted)}, thermal \textbf{(dashed)}, and magnetic \textbf{(solid)} energies, integrated over the whole simulation domain. The magnetic energy grows several orders of magnitude during the amplification phase, reaching a saturation value that depends only moderately on the BH spin. 
		\label{energies}}
\end{figure}

All the binaries, differing only by the black hole's spin $a/M_{\rm{BH}}=(0.2,0.35,0.5)$, perform almost two orbits before the neutron star is tidally disrupted by the BH. The qualitative dynamics afterwards can be followed in Fig.~\ref{RHOF}, which displays time snapshots of the density in the equatorial plane for the three spin cases. Most of the NS is swallowed by the BH in a short time scale of a few milliseconds, except the matter located more distant from the BH, outside the ISCO (i.e., $r \geq R_\mathrm{ISCO}$), which forms a one-arm spiral structure. Since $R_\mathrm{ISCO}$ decreases as the BH spin increases, this one-arm spiral is more massive for high spins.
The fluid in the outermost region becomes unbounded and leads to the dynamical ejecta. The bound component winds around the BH, eventually interacting with its own tidal tail. 

The Kelvin-Helmholtz instability~(KHI) develops in the contact interfaces of the one-arm spiral colliding with itself, forming a turbulent, compact accretion disk with the shape of a torus in $t \lesssim\,10\,$ms. During this stage, the magnetic field is continuously twisted by the fluid motion, efficiently converting kinetic into magnetic energy. 
As a result, the turbulent dynamo induces exponential growth of the magnetic field, until roughly the time of formation of an almost axisymmetric, rapidly rotating accretion disk at $t \lesssim\,2\,$ms.
Once the shear layers have been dissipated and the KHI is no longer active, the turbulence slowly decays in the disk. At this point  other instabilities and amplification mechanisms, such as magnetic winding and the magneto-rotational instability~(MRI), might affect and even dominate the magnetic field dynamics.

Several relevant quantities, namely the density, the poloidal and toroidal components of the magnetic field, and the inverse of the $\beta$-parameter, are displayed on a meridional plane at $t=20\,$ms in Fig.~\ref{meridional}. Here it is clear that the BH spin strongly determines how much mass is swallowed by the BH and how much remains in the accretion disk after the star's disruption. The poloidal and toroidal components of the magnetic field are comparable in magnitude and display a rapid time and spatial variation, indicating a high degree of turbulence. We can also observe that the magnetic field is stronger for higher spins.  However, the inverse of the plasma parameter $\beta$ is smaller at higher spins, basically due to the lower densities and pressures in the tenuous disk produced by small spins.

At late times, the disk becomes more axisymmetric, as can be seen in the bottom row of Fig.~\ref{RHOF} as well as the near symmetry shown in the meridional profiles of the disk shown in Fig.~\ref{meridional}.
The disk has a torus-like shape with three distinct regions: a high density, compact region forming the \textit{bulk} of the disk (roughly the region containing $75\%$ of the disk's mass), an extended corona or \textit{envelope} with low-density fluid, and then a very rarefied cone-shaped \textit{funnel} defined where the density is below $10^8 \rm{g/cm}^3$. The funnel extends roughly $45^o$ around the vertical $z$-axis for the smallest spin and approximately $35^o$ for the highest one.
The magnetic field is strongest in the innermost part of the disk (i.e., close to the BH), mainly due to the flux freezing condition of MHD; magnetic flux is frozen with the fluid, and so as the fluid advects toward the BH, the magnetic field grows in strength. Consistent with this explanation, the inverse of the beta-parameter is fairly uniform within the disk.

\subsection{Global quantities and surface fluxes}
\begin{figure}[h]
	\includegraphics[width=8.5cm]{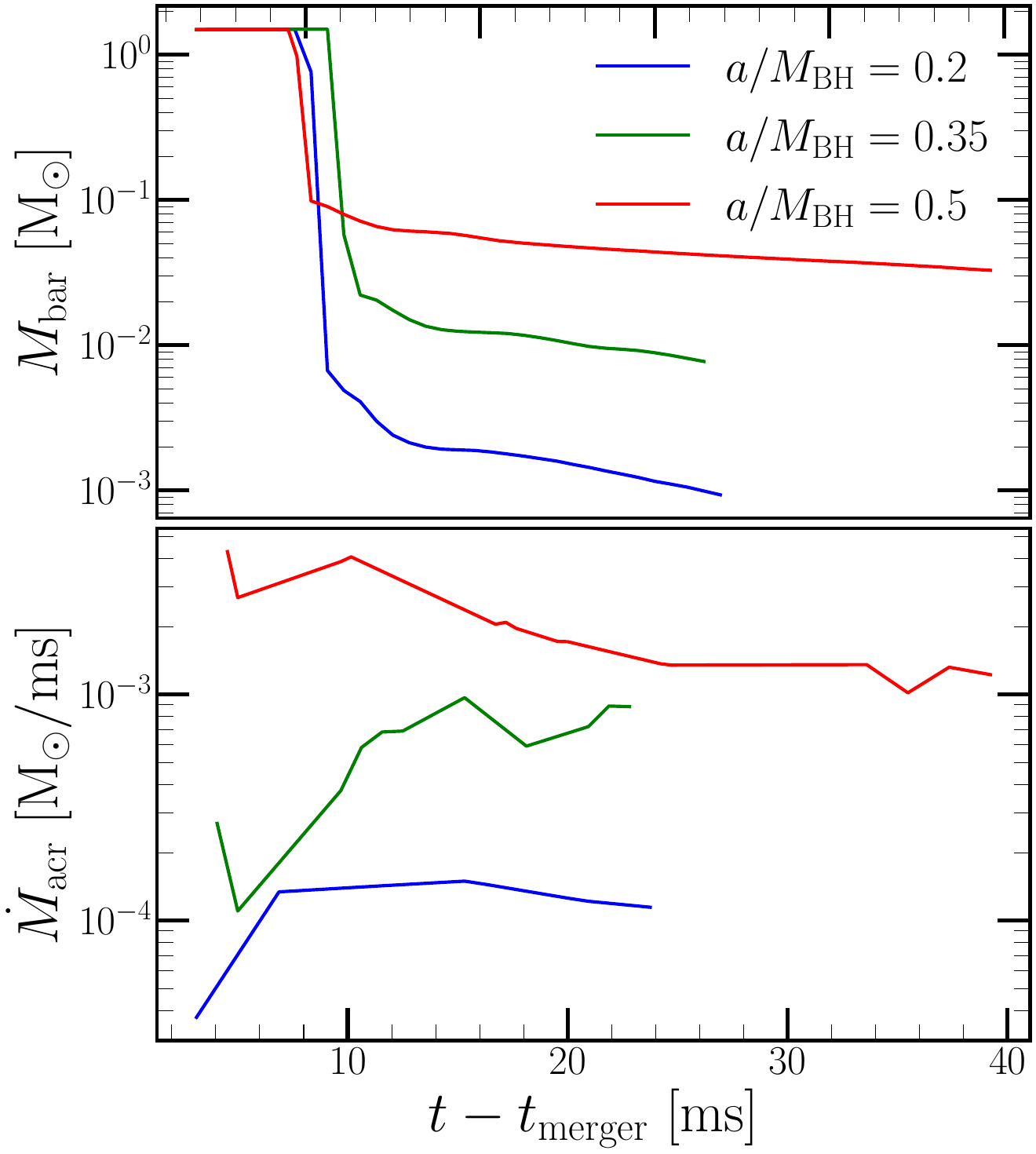}
	\caption{\emph{Mass evolution.}
    Baryonic mass of the system, integrated over the whole simulation domain \textbf{(top)}. Massive disks are only produced for high spins $a/M_{\rm{BH}} \gtrsim 0.5$. Mass accretion rate on the BH \textbf{(bottom)}. 
		\label{all_masses}}
\end{figure}

\begin{figure}[h]
	\includegraphics[width=8.5cm]{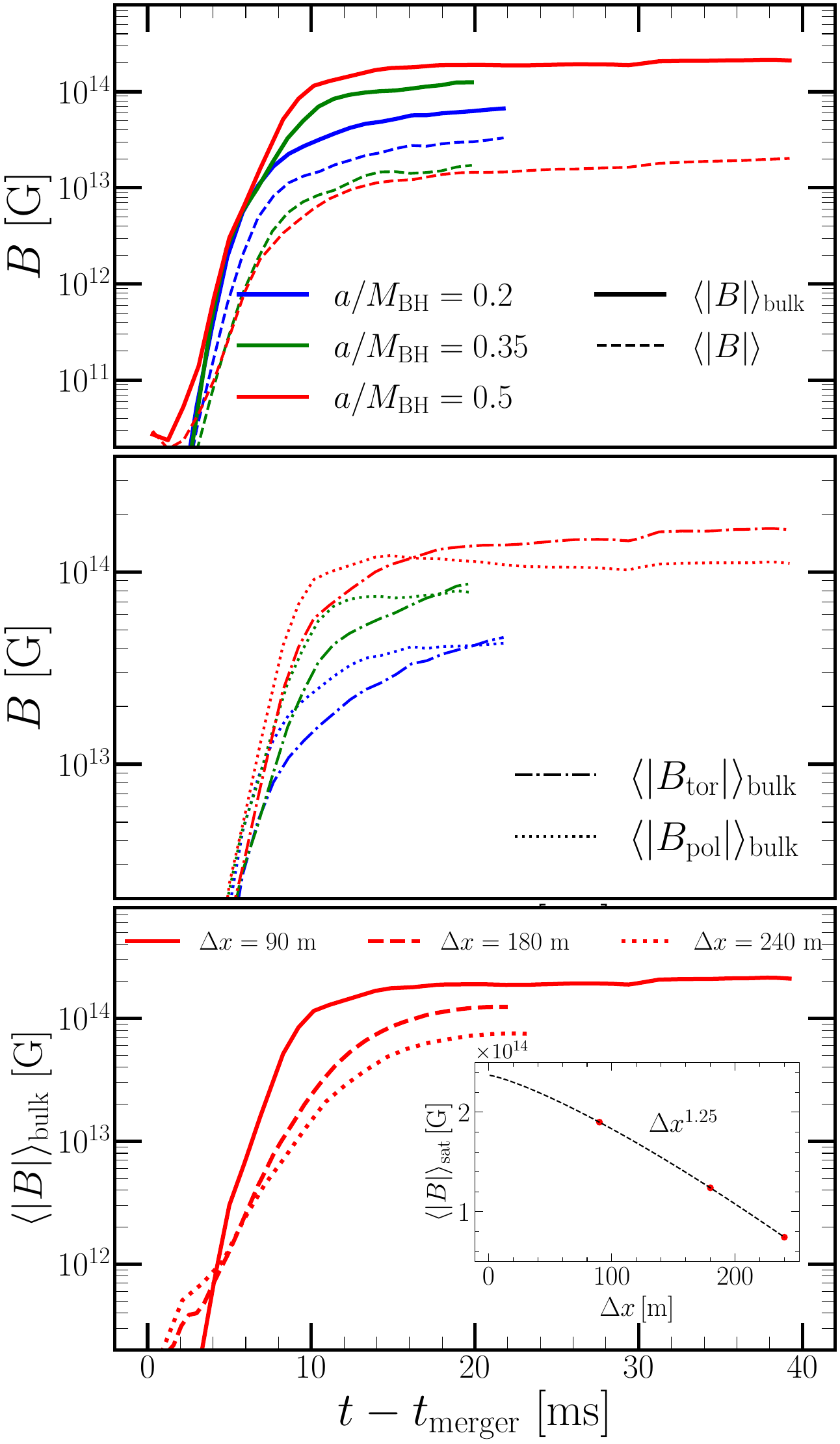}
	\caption{\emph{ Magnetic field evolution.}
Evolution of the strength of the magnetic field, averaged over a region containing either only the bulk or the whole the disk remnant \textbf{(top)}. The magnetic field in the bulk is amplified to values $\approx 10^{14}\, \text{G}$ for all the BH spins. Poloidal and toroidal components of the magnetic field, averaged in the bulk \textbf{(middle)}. They remain comparable during the expontial amplification phase, suggesting that it is due to isotropic, turbulent small-scale dynamo. Notice that the averaged toroidal component grows roughly linearly after $t \gtrsim 20\,\text{ms}$. Strength of the magnetic field in the bulk for different spatial resolutions, showing roughly linear convergence of the saturation values with resolution \textbf{(bottom)}.
		\label{Baverages}}
\end{figure}

\begin{figure}[h]
	\includegraphics[width=8.5cm]{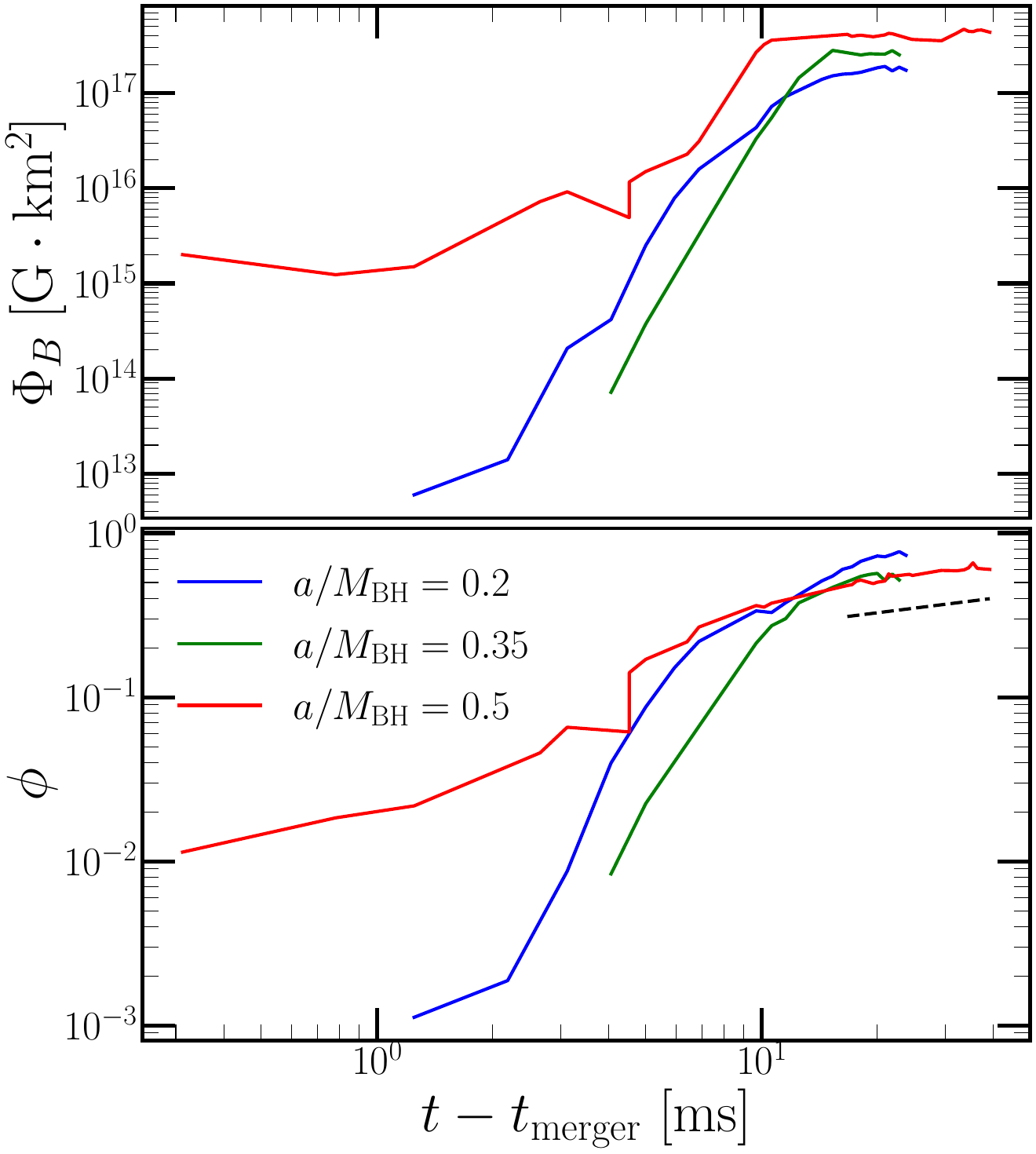}
	\caption{\emph{Magnetic flux.} 
Evolution of the magnetic flux \textbf{(top)} and the dimensionless magnetic flux \textbf{(bottom)}, evaluated near the BH horizon. The magnetic flux quickly grows until approximately $t \approx 15$~ms and then remains mostly constant. The dimensionless magnetic flux reaches values $\phi \lesssim 1$ at $t \approx 20$~ms and then keeps growing at a slower rate \textbf{(dashed)}. 
		\label{MAD}}
\end{figure}

A more quantitative analysis can be obtained by integrating relevant quantities either over the whole domain or in specific regions. 
In the top panel of Fig.~\ref{energies} the rotational, thermal, and magnetic energies are displayed as functions of time from the merger. Note that the magnetic energy is initially very small, corresponding to the realistic magnetic fields $|B| \leq 10^{11}\,\text{G}$. Despite beginning with low values of roughly $10^{44}$~ergs for $a/M_{\rm{BH}}=0.5$ and below $10^{40}$~ergs for the other spins~\footnote{The specific value depends on the origin of the disrupted matter that is finally forming the disk; more massive disks are produced from inner shells that were more strongly magnetized.}, the magnetic energy grows beyond $10^{48}\, \rm{ergs}$ in less than $20\rm{ms}$ due to the turbulent small-scale dynamo. 

The disk's mass depends very strongly on the BH spin (i.e., roughly $M_\mathrm{disk} \propto \left(a/M_{\rm{BH}}\right)^5$) for the explored BH spin range, with the mass ratio and equation of state employed here. As expected, the rotational and thermal energies depend almost linearly on the remnant disk mass 
($E_\mathrm{rot} \propto \left(M_\mathrm{disk}\right)^{0.9}$, $E_\mathrm{th} \propto \left(M_\mathrm{disk}\right)^{0.85}$), since the angular velocity in the disk and the fraction that is converted into heat does not change significantly for the three configurations.
However, the magnetic energy dependence on the disk's mass is milder, $E_\mathrm{mag} \propto \left(M_\mathrm{disk}\right)^{0.7}$, reaching saturation values of ${\cal O}(10^{47}-10^{48})$~ergs for our range of BH spins. 
Even with a small disk mass, the turbulent dynamo actively  amplifies the magnetic field energy up to relatively high values.

The evolution of mass in different regions is further analyzed in Fig.~\ref{all_masses}. 
The total baryonic mass is displayed in the top panel, showing the initial mass of the neutron star, the sharp decrease when the disrupted matter infalls into the black hole, and finally the mass remaining in the accretion disk. We can observe that the disk mass spans almost two orders of magnitude for the range of spins considered here. 
The mass accretion rate onto the BH is displayed in the bottom panel, showing that the accretion rate is smaller for smaller disk mass.

The magnetic field evolution can be further analyzed by computing its components and average magnitude components both in the bulk and in the whole disk~\footnote{We have also computed these averages by weighting the respective quantities by density as is often done in the literature, and the results after disk formation are quite similar to those obtained via our averaging over the bulk.}, as displayed in Fig.~\ref{Baverages}.
For all our spins, the magnetic field strength reaches approximate values of ${\cal O} (10^{13}\,\text{G}) $ when it is averaged in the whole disk, and ${\cal O} (10^{14}\,\text{G})$ when averaged only in the bulk. The poloidal and toroidal averages are comparable and grow at similar rates during the strongest amplification early phase $t \lesssim 20\,\text{ms}$ for all the cases. This equipartition between the two magnetic field components is another indication that isotropic turbulence is driving the magnetic field amplification
during this stage of the disk formation. 

After this period, the toroidal component starts growing almost linearly with time while the poloidal component remains roughly constant, suggesting that winding is taking over as the dominant amplification mechanism. Finally, the bottom panel displays the averaged magnetic field in the bulk for different resolutions. 
The difference between the saturation values obtained with the resolutions $\Delta x =(180, 90)\,$m are much smaller than the one obtained with the resolution $\Delta x =(180, 240)\,$m.
In particular, the inset suggests that the saturation magnetic field value converges slightly better than linearly. Given the turbulent nature of the bulk fluid and the fact that high resolution shock capturing methods reduce to first order accuracy at shocks, such a low order of convergence is not surprising. Our extrapolation, assuming an error proportional to $\Delta x^{1.25}$, suggests that the true saturation value is within 30\% of our most resolved case.

The evolution of the magnetic flux  on the BH is plotted in the top panel of Fig.~\ref{MAD}, with the bottom panel displaying the dimensionless magnetic flux. Several interesting features can be observed in these quantities. First, similar to the magnetic field evolution, there is significant growth during the first stage of accretion. Both magnetic fluxes quickly saturate at approximately $20~\textrm{ms}$ and then their evolution slows significantly. The dimensionless magnetic fluxes begin to grow slower when reaching $\phi \lesssim 1$ for all spins, much smaller than the saturation value of $ 50 $ necessary for magnetically arrested disks (MADs)\citep{Tchekhovskoy2011}. If the magnetic flux $\Phi_B$ remains nearly constant and the mass accretion rate decays after a transitional phase lasting $ t \lesssim 100\,{\rm ms}$ with the expected power-law $\dot{M}_{\rm{disk}} \propto t^{-\zeta}$ with $ \zeta \approx 1.5-2 $ found in long-term simulations~\cite{2022PhRvD.106b3008H,Hayashi:2022cdq,Gottlieb2023,Gottlieb2023b}, it would result in $\phi \sim \Phi_B \dot{M}_{\rm{disk}}^{-1/2} \propto t^\xi $, where $ \xi \approx 0.75-1 $. Indeed, these previous simulations have shown that $ \phi $ grows at most linearly with time, exhibiting similar normalization values at $ \sim 20\,{\rm ms} $ as found here. This result might have important implications for the jet duration, as discussed in Sec.~\ref{sec:discussion}. The precise values of the energies, masses, magnetic field, and fluxes at saturation can be found in Table~\ref{tab:spin_results}. 

\begin{figure}[t!]
	\includegraphics[width=8cm]{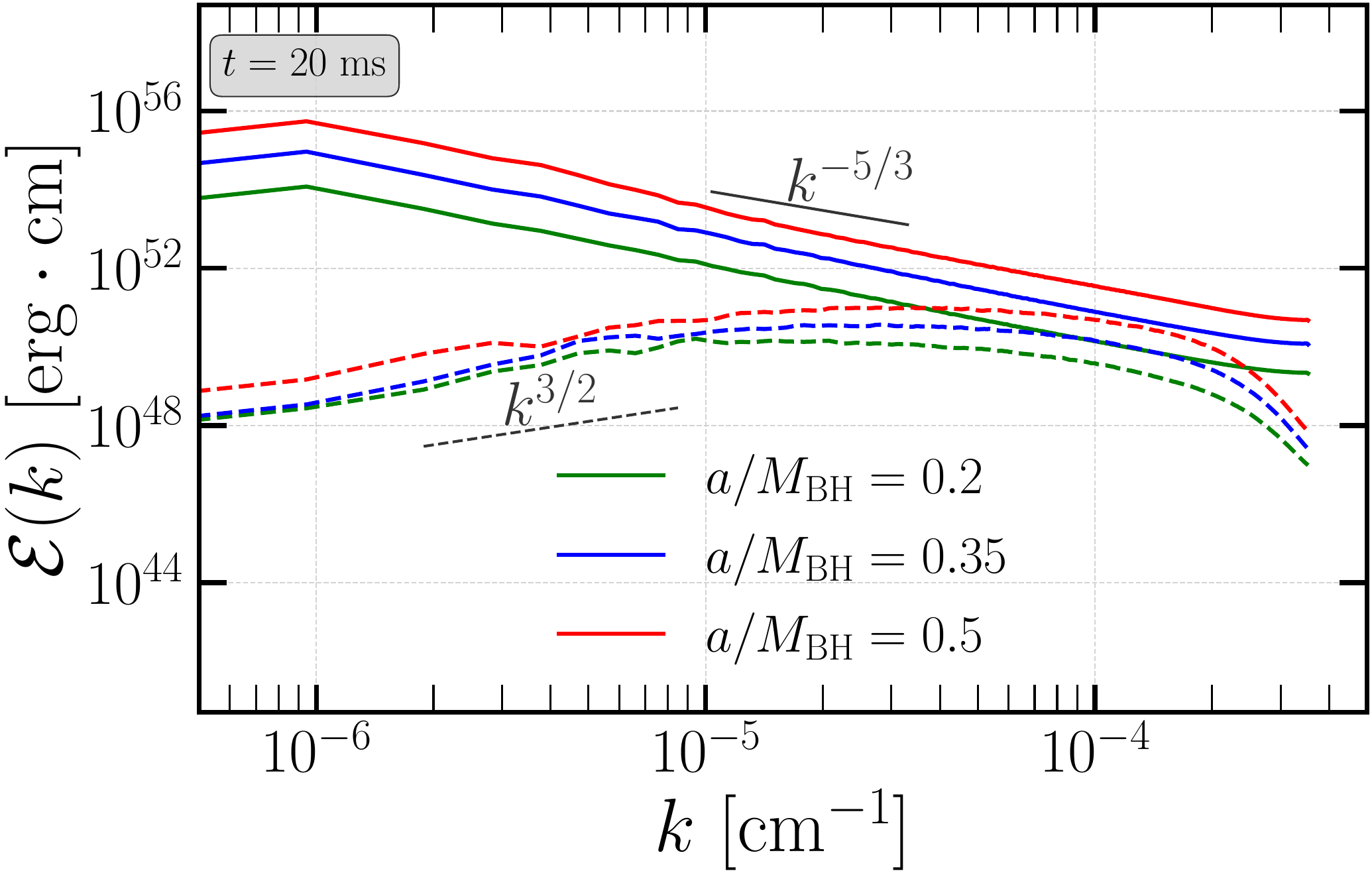}
	\caption{\emph{Energy spectra.} Kinetic \textbf{(solid)} and magnetic \textbf{(dashed)} spectra for the three spin cases at $t = 20\,\rm{ms}$. The kinetic energy spectra show the standard Kolmogorov power law $k^{-5/3}$ (short solid blue line) in the inertial range. The magnetic energy spectra follow the Kazantsev power law $k^{3/2}$ \textbf{(short dashed blue line)} at low wavenumbers. 
		\label{spectra_time}}
\end{figure}

\begin{figure*}[tb] %
	\centering
	\includegraphics[width=\textwidth]{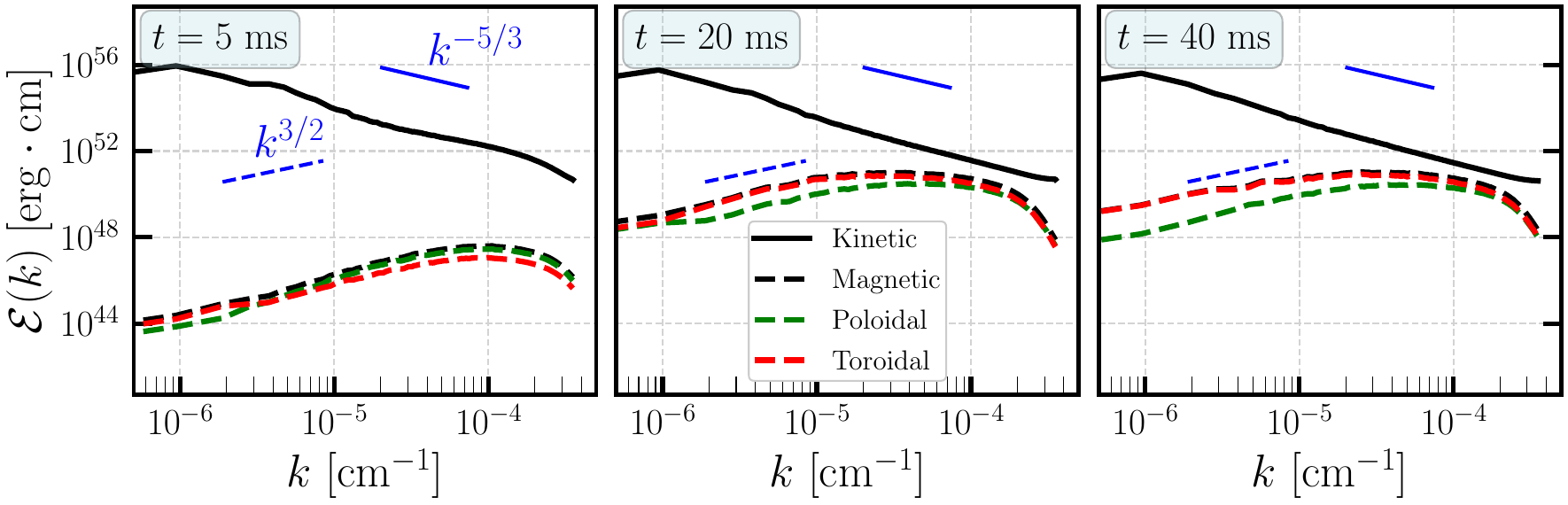}
	\caption{\emph{Energy spectra.} Kinetic \textbf{(solid)} and magnetic \textbf{(dashed)} spectra for the case with $a/M_{\rm{BH}}=0.5$ at $t = (5, 20, 40) \,\rm{ms}$. The magnetic energy grows quickly in the small scales until $t \sim 20$~ms, and then slowly in the large scales (i.e., deviating from the Kazantsev power law) afterwards.
		\label{spectra_time_a05}}
\end{figure*}

\subsection{Spectra}
A standard approach in problems displaying turbulent dynamics involves the analysis of the energy spectra.
The kinetic and magnetic energy spectra are displayed at $t=20$~ms in Fig.~\ref{spectra_time} for the three different spins.
Both sets of spectra follow the behavior expected in a turbulent, weakly magnetized fluid: the kinetic energy spectrum decreases with the \textit{Kolmogorov power law dependence} $k^{-5/3}$ while the magnetic field spectrum at low wave-numbers follows the \textit{Kazantsev power law} $k^{3/2}$~\cite{1968JETP...26.1031K,1992ApJ...396..606K}. 
We observe that the kinetic spectra for different spins are basically the same, up to a scale factor. The magnetic field spectra also have a similar shape, although the peak is flattened for lower BH spins. 

The evolution of the spectra for $a/M_{\rm{BH}}=0.5$ is displayed in Fig.~\ref{spectra_time_a05}. The initial magnetic field before the disruption is mainly poloidal and confined within the star, yielding a bump in the spectra centered at $k \approx 10^{-5} \textrm{cm}^{-1}$. During disk formation, the transfer from kinetic to magnetic energy occurs at all scales (the magnetic spectra move up), but it is clearly more effective at high wavenumbers (i.e., small scales) as this bump moves towards higher wavenumbers. This behaviour, combined with the isotropy (i.e., keeping similar values of the poloidal and toroidal components) is expected in a turbulent regime.
After $20$~ms the growth occurs mostly in the toroidal component, at small wavenumbers, due mainly to two effects driven by the winding: a transfer from kinetic to magnetic energy, and a small conversion from poloidal into toroidal magnetic field. 

\begin{figure}[h]
	\includegraphics[width=8.5cm]{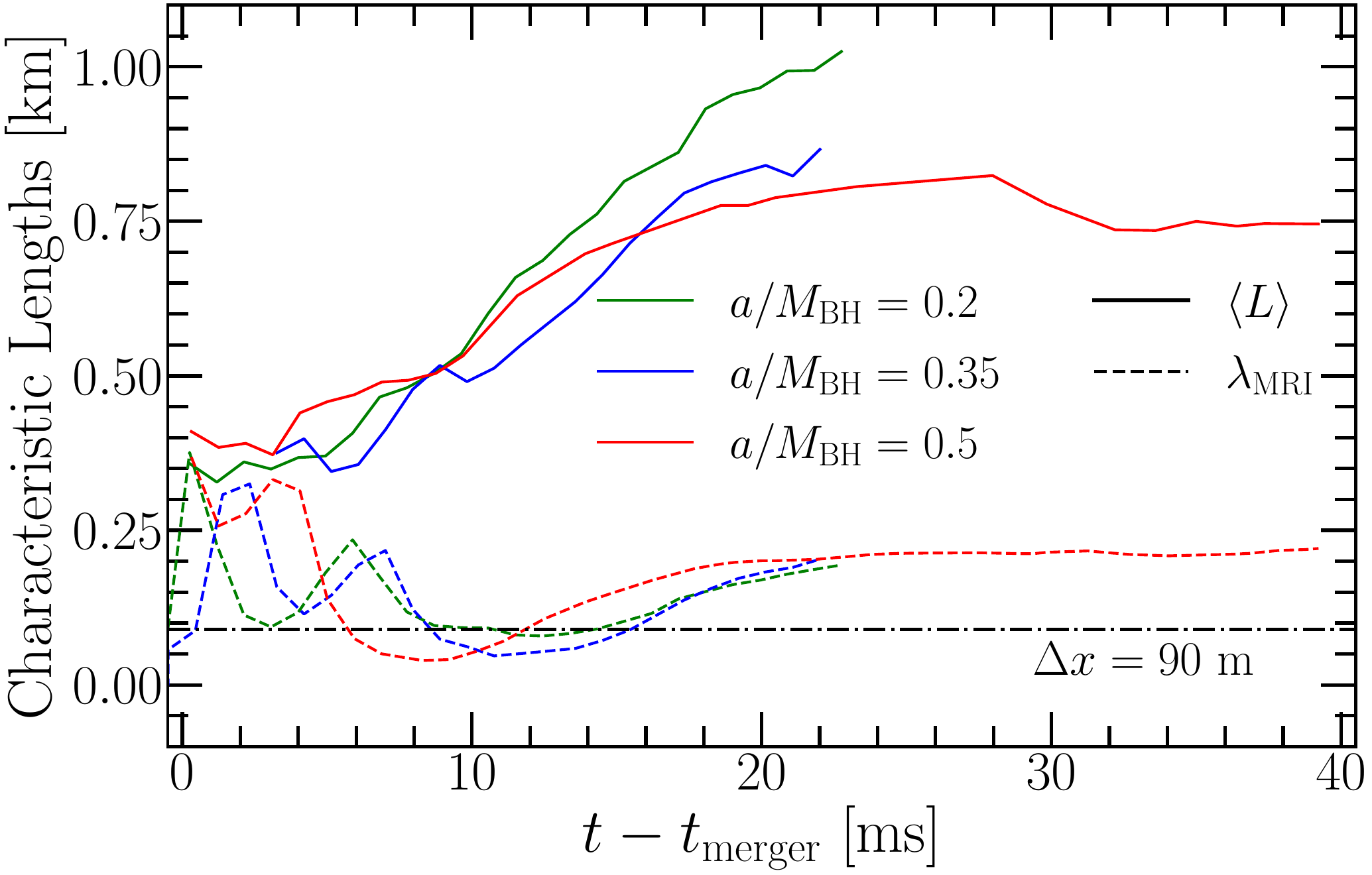}
	\caption{\emph{Characteristic length of the magnetic field.} The time evolution of the characteristic coherent length of the magnetic field defined by Eq.~\eqref{eq:coherence}.  After disk formation, the coherence scale grows monotonically with time, slowing down significantly (or even saturating) for $t \gtrsim 20\,\text{ms}$. The wavelength of the fastest growing mode of the MRI is much smaller than the coherent scale, but comparable to the minimum resolution in this region. \label{characteristic_scale}
	}
\end{figure}

Finally, the time evolution of the coherence length, see Eq.~\eqref{eq:coherence},  of the magnetic field is displayed in Fig.~\ref{characteristic_scale} for all cases. It increases monotonically until $t\approx 20$~ms with comparable values across disk masses, though slightly higher at lower spins. Although we have not followed the evolution of lower spin cases far enough to confirm saturation, it is reasonable to expect a qualitative behavior similar to the highest spin case. 

In Fig.~\ref{characteristic_scale} we compare this coherence length with both the wavelength of the fastest growing mode of the MRI and the finest spatial resolution $\Delta x$. Physically, as argued in Ref.~\cite{Palenzuela:2021gdo}, the development of the MRI requires a sufficiently large scale magnetic field. Here that condition can be expressed as $\langle L \rangle \gg \lambda_\mathrm{MRI}$, which is marginally satisfied once the magnetic field reaches saturation at $t\approx 20\,$ms. Numerically, the MRI can only be accurately resolved when its fastest growing mode is captured with a sufficient number of grid points, namely $\lambda_\mathrm{MRI}\gg \Delta x$. The figure shows that this condition is not satisfied, indicating that we are still far from resolving the MRI in these systems. 

%
\subsection{Radial dependence}

\begin{figure*}[t]
	\includegraphics[width=15cm]{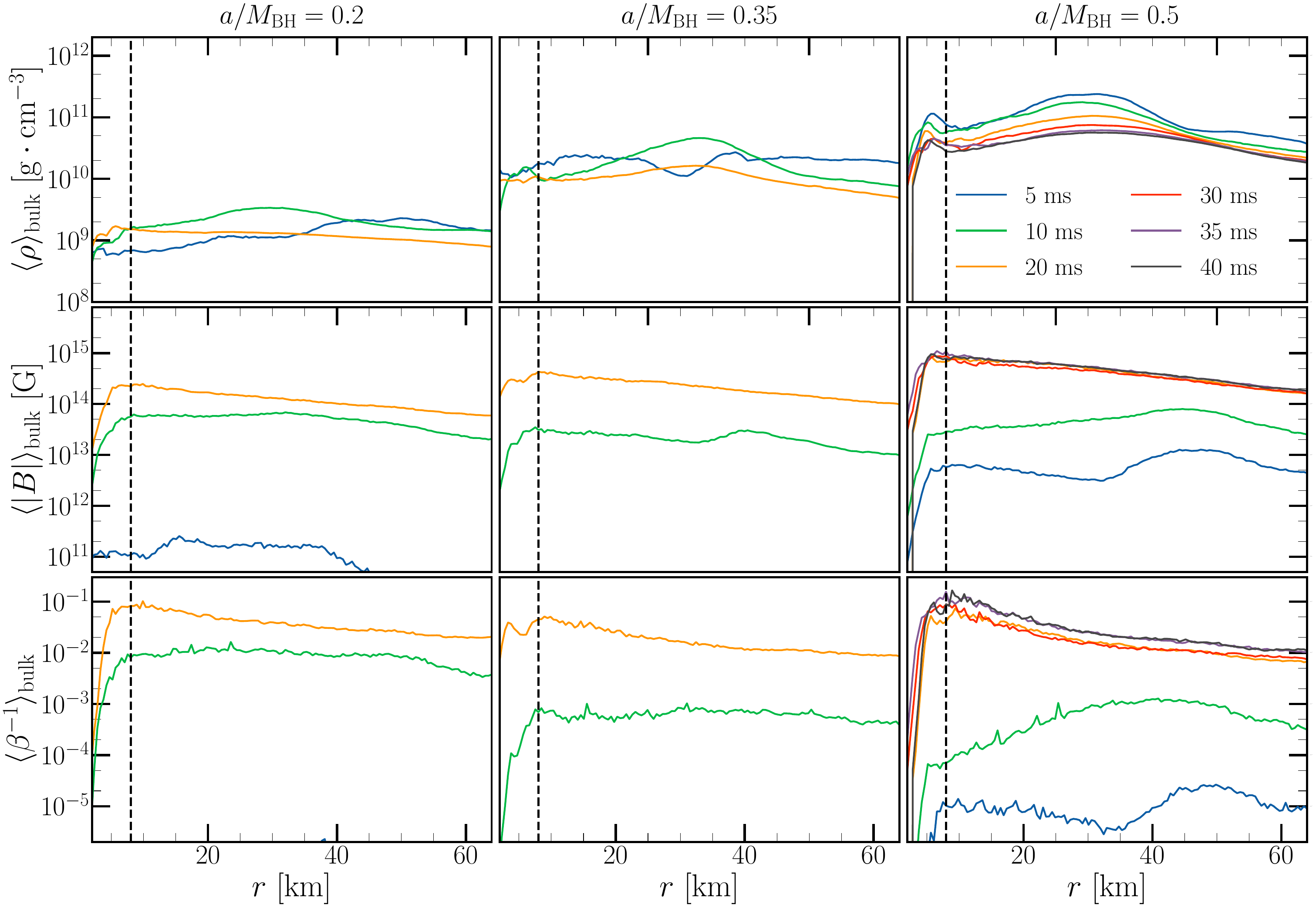}
	\caption{\emph{Cylindrical disk averages.}
		The rest-mass density \textbf{(top)}, the magnetic field \textbf{(middle)}, and the ratio $\beta^{-1}$ \textbf{(bottom)} are displayed as functions of radii at different times. The average magnetic field strength increases rapidly from $10^{11}\, \text{G}$ to more than $10^{14}\, \text{G}$ in about $20 \, \text{ms}$. The growth is similar for all radii at early times but is mostly suppressed at late time $t \ge 20 \, \text{ms}$ except at the innermost region. The BH apparent horizon is depicted as a vertical dashed line. Note that, at early times and for some of the low spin cases, the magnetic field and the ratio $\beta^{-1}$ are  not visible because they are smaller than the vertical ranges.
		\label{several_cylinders_times}}
\end{figure*}

To understand the radial structure of the accretion disk as a function of the BH spin, Fig.~\ref{several_cylinders_times} depicts the radial profiles of the cylindrical averages of the density, magnetic field, and the inverse of the beta-parameter.

The density profile shows a clear bump at late times $t \gtrsim 20 \, \rm{ms}$ for the spin $a/M_{\rm{BH}}=0.5$. The maximum of the bump which moves inwards, towards the horizon, as the BH spin decreases. These results confirm that not only the mass of the disk, but also its shape depends strongly on the spin.
The magnetic field evolution is displayed in the middle panel, showing a rapid growth throughout all the domain, but especially in the innermost part of the disk. At late times $t \gtrsim 20\,\text{ms}$, the growth rate is almost completely suppressed. 

Interestingly, in this quasi-stationary state, the magnetic field has a similar shape for all spins, with a maximum near the horizon and then decaying radially. The inverse of the beta-parameter increases quickly in the first $20$ms after the merger, reaching peak values of ${\cal O}(0.1)$ for the lowest spin $a/M_{\rm{BH}}=0.2$ and decreasing slightly as the spin increases. In any case, even for this case, the fluid pressure dominates at all radii for the duration of our simulations.

\section{Discussion}\label{sec:discussion}

We have performed numerical simulations modeling the coalescence of a black hole with a neutron star to study the magnetic field dynamics occurring after the merger, extending and improving on our recent preliminary results~\cite{Izquierdo:2024rbb}.
We have considered systems with a fixed mass ratio $q=3$ but with BH spin spanning the range $a/M_{\rm{BH}}=(0.2, 0.35, 0.5)$, such that the NS tidally disrupts and produces accretion disks with masses in the range $M_{\mathrm{disk}} \approx 0.001-0.1 M_{\odot}$.
We initialize the NS magnetic field with an average strength of $10^{11}\,\rm{G}$, a realistic upper bound for Gyr-old NS, similar to the ones commonly found in binary systems. This field represents a tiny fraction of the total energy of the system and is much smaller than commonly used in current numerical simulations.
Using high-order schemes, LES techniques with the gradient SGS model, and the highest resolution of this type of simulation $\Delta x \lesssim {\cal O}(100\,\rm{m})$, we resolved the saturation of the magnetic field resulting from the turbulent dynamics during the disk formation. 

Our simulations show that, shortly after tidal disruption, the magnetic field is first amplified by the KHI. A shear layer in the one-arm spiral, as it interacts with itself while being wound around the final black hole, is prone to unstable modes that develop vortices at all scales. Turbulent dynamics, driven mainly by this instability, amplifies the magnetic field, which is then redistributed throughout the torus by the fluid flow. A quasi-stationary accretion disk, maintaining some degree of small-scale turbulence, is achieved after approximately $10\,\rm{ms}$. The average magnetic field strength at the saturation phase is approximately $10^{14}\,\rm{G}$. As expected from isotropic turbulence, the poloidal and toroidal components of the magnetic field have comparable strengths at saturation. At later times, the magnetic field grows linearly due to the winding mechanism. 
Eventually, small-scale dynamo induced by the MRI is expected to dominate the magnetic field amplification, although our results indicate that even higher resolutions might be needed to capture this instability.

We examine in detail the magnetic flux evolution, and argue that our results, along with other work, support the claim that all BH-NS mergers produce a new subpopulation of long GRBs (lbGRBs). Our simulations show that the evolution of the dimensionless magnetic flux threading the BH is roughly the same across a wide range of disk masses. Assuming that this quasi-universal evolution holds across various equations of state and mass ratios, we can estimate the time required for the system to reach a MAD state ($\phi \approx 50$).
In particular, beginning with values we obtain at $t=20\, \mathrm{ms}$ and assuming $\phi \propto t^\xi $ with $ \xi \approx 0.75-1 $ as found in other, longer duration numerical simulations (i.e., see for example ~\cite{2022PhRvD.106b3008H,Hayashi:2022cdq,Gottlieb2023,Gottlieb2023b}), one expects a MAD state at $t \gtrsim 10\,\mathrm{s}$~\citep{Gottlieb2023}.
Since the transition to MAD marks the end of the prompt GRB emission \citep{Gottlieb2023}, the universality of this transition time for all BH--NS mergers would indicate that all BH--NS mergers produce long-duration GRBs. However, since the disk mass determines the resulting jet power, only massive disks would produce jets bright enough to be observable~\citep{Gottlieb2024}. Our findings suggest that BH--NS mergers contribute exclusively to long-duration GRBs \citep{Gottlieb2024}, leaving NS--NS mergers as the most likely central engines of standard short-duration GRBs \citep{Gottlieb2024}.

\subsection*{Acknowledgements}

This work was supported by the Grant PID2022-138963NB-I00 funded by MCIN/AEI/10.13039/501100011033/FEDER, UE. This work was also supported by the National Science Foundation via grants PHY-2308861 and PHY-2409407.MRI thanks financial support PRE2020-094166 by MCIN/AEI/PID2019-110301GB-I00 and by “FSE invierte en tu futuro”. OG is supported by the Flatiron Research Fellowship. The Flatiron Institute is supported by the Simons Foundation. MB acknowledge partial support from the STFC Consolidated Grant no. ST/Z000424/1. The authors thankfully acknowledges RES resources provided by BSC in MareNostrum to AECT-2024-2-0004 and AECT-2024-3-0007. This work used the DiRAC@Durham facility managed by the Institute for Computational Cosmology on behalf of the STFC DiRAC HPC Facility (www.dirac.ac.uk). The equipment was funded by BEIS capital funding via STFC capital grants ST/P002293/1, ST/R002371/1 and ST/S002502/1, Durham University and STFC operations grant ST/R000832/1. DiRAC is part of the National e-Infrastructure. This research used Frontera at the Texas Advanced Computing Center, made possible by National Science Foundation award OAC-1818253, and Expanse via the Advanced Cyberinfrastructure Coordination Ecosystem: Services \& Support (ACCESS) program, which is supported by National Science Foundation grants 2138259, 2138286, 2138307, 2137603, and 2138296.

\bibliographystyle{utphys}
\bibliography{biblio}

\end{document}